\documentclass[prd,aps,floats,superscriptaddress,nofootinbib,showpacs,twocolumn,floatfix]{revtex4}
\usepackage{exscale}                  
\usepackage[intlimits]{amsmath}       
\usepackage{amsfonts}
\usepackage{amssymb,amscd}
\usepackage{epsfig}
\usepackage{wrapfig}

\newcommand{\GS}[1]{#1\!\!\!\!\!\!\!\not~}	
\newcommand{\Dslash}{\GS{D}}

\newcommand{\pslash}{p\!\!\!/\,}

\newcommand{\cO}{\mathcal{O}}

\begin{document}

\title{Finite volume effects and dynamical chiral symmetry breaking 
in $\mbox{QED}_3$}

\author{Tobias Goecke}
\affiliation{Institute for Nuclear Physics, 
 Darmstadt University of Technology, 
 Schlossgartenstra{\ss}e 9, 64289 Darmstadt, Germany}
\author{Christian~S.~Fischer}
\affiliation{Institute for Nuclear Physics, 
 Darmstadt University of Technology, 
 Schlossgartenstra{\ss}e 9, 64289 Darmstadt, Germany}
\affiliation{GSI 
Helmholtzzentrum f\"ur Schwerionenforschung GmbH, 
  Planckstr. 1,  64291 Darmstadt, Germany.}
\author{Richard Williams}
\affiliation{Institute for Nuclear Physics, 
 Darmstadt University of Technology, 
 Schlossgartenstra{\ss}e 9, 64289 Darmstadt, Germany}

\begin{abstract}
We investigate the impact of finite volume effects on the critical
number of flavours, $N_f^c$, for chiral symmetry restoration in 
$\mbox{QED}_3$. To this end we solve a set of coupled Dyson-Schwinger 
equations on a torus. For order parameters such as the anomalous dimension 
of the fermion wave function or the chiral condensate we find substantial 
evidence for a large dependence on the volume. We observe a shift in 
$N_f^c$ from values in the range of $3.61 \le N_f^c \le 3.84$ in the 
infinite volume/continuum limit down to values below $N_f \le 1.5$ at 
finite volumes in agreement with earlier results of Gusynin and Reenders 
in a simpler truncation scheme. These findings explain discrepancies in 
$N_f^c$ between continuum and lattice studies. 
\end{abstract}

\maketitle

\section{Introduction}

Dynamical chiral symmetry breaking in the theory of Quantum 
Electrodynamics in (2+1) dimensions, QED$_3$, \cite{Pisarski:1984dj} has been 
studied quite intensively over the years. The problem is of considerable 
interest for two reasons. On the one hand, QED$_3$ has enough similarities to 
Quantum Chromodynamics (QCD) such that analogies to the more complicated 
non-Abelian case may be drawn. On the other hand, QED$_3$ itself is of 
considerable interest due to possible applications in condensed matter systems. 

In particular, QED$_3$ with
$N_f=2$ flavours of massless fermions has been suggested as an effective
low energy theory of high-$T_c$ cuprate superconductors 
\cite{Herbut:2002yq,Franz:2002qy,Nersesyan}. These possess an unconventional 
$d$-wave symmetry of the pairing condensate, with nodes at the electronic
Fermi surface allowing for a description in terms of massless nodal 
quasi-particles. The quasi-particles do not couple to an external electromagnetic field
and hence represent pure spin degrees of freedom; a spin-charge separation
has taken place. The interaction of these spinons with the collective, 
topological excitations of the gap can be described by a $U(1)$ gauge theory. 
Since furthermore the motion of the quasi-particles is mainly confined to 
the two-dimensional copper-oxygen planes in these systems one ends up with 
quantum electrodynamics in (2+1) dimensions. Depending on whether the system 
is ordered or disordered it is either in an insulating quantum antiferromagnetic 
(AF) phase or a pseudogap (PG) phase with remnant properties of the underlying 
superconductor. In the gauge theory the AF-phase corresponds to a phase with 
broken chiral symmetry and long-range correlations due to massless photons.
In the PG-phase the fermions are massless and the fermion wave function as well 
as the photon propagator develop power laws at small momenta with a 
fractional anomalous dimension \cite{Franz:2002qy,Fischer:2004nq}. 

The above considerations explain the interest in determining $N_f^c$, the
critical number of fermion flavours for the chiral phase transition of QED$_3$.
If $N_f^c > 2$ then the effective low energy theory is chirally broken 
at zero temperature. For the superconductor this means that the theory displays
a phase transition between the superconducting and the antiferromagnetic phase
when doping is varied \cite{Herbut:2002yq,Franz:2002qy}. If on the other hand 
$N_f^c < 2$ the system goes from a superconducting to a pseudogap-phase 
when underdoped.

The value of $N_f^c$ has been 
investigated in a number of studies from Dyson-Schwinger equations (DSEs) 
in the continuum \cite{Franz:2002qy,Fischer:2004nq,Appelquist:1986fd,
Appelquist:1988sr,Pennington:1988jw,Atkinson:1989fp,Pennington:1990bx,Burden:mg,
Maris:1995ns,Gusynin:1995bb,Maris:1996zg} and lattice gauge theory 
\cite{Dagotto:1988id,Hands:1989mv,Hands:2002dv,Hands:2004bh,Strouthos:2008hs}. 
While the former 
allows for numerical as well as analytical studies in principle there remains 
the question of the importance of truncation artefacts. For $N_f^c$ a partial
answer has been obtained in \cite{Fischer:2004nq}, where it was found that the 
details of the fermion-photon vertex only have a minor quantitative impact 
on $N_f^c$. In particular for all vertex dressings employed $N_f^c$ stayed 
well above $N_f = 2$. 

In all studies of dynamical chiral symmetry breaking in continuum QED$_3$ a clear
separation of scales has been found. The intrinsic scale of the theory is 
given by the dimensionful coupling constant $\alpha = N_f \, e^2/8$. Then a 
second and much lower scale is given by the dynamically generated fermion mass 
$M(0)$ in the chiral limit. Related to this one has small values for the other
order parameter, the chiral condensate. On the lattice with its finite volume
this separation of scales is hard to bridge. Consequently, recent studies for
the number of flavours $N_f=2$ \cite{Hands:2002dv} and $N_f=4$
\cite{Hands:2002dv,Hands:2004bh} determined bounds on the chiral
condensate, but no definite value for $N_f^{\mathrm crit}$ could be
extracted. A definite signal for chiral symmetry breaking was obtained only 
for $N_f=1$ \cite{Hands:2004bh}. A very recent calculation claims that 
$N_f^c \approx 1.5$ \cite{Strouthos:2008hs} with the caveat of volume and
discretisation artefacts.

In general, the presence of an infrared cutoff due to the finite volume reduces the 
value of the critical number of flavours. This has been demonstrated by Gusynin 
and Reenders in a simple truncation scheme for the DSEs \cite{Gusynin:2003ww}.
They considered an approximation to the DSE for the fermion self-energy which 
neglects corrections to the fermion wave-function and the fermion-photon vertex.
The photon is then given by its leading behaviour in the $1/N_f$ expansion.  
In this work we elaborate on these findings by considering a more sophisticated
truncation scheme which explicitly takes into account non-perturbative effects
in all of these quantities. In addition we follow a different strategy to assess
the volume effects by evaluating the system on a three-torus\footnote{A 
corresponding technique has been applied in QCD$_4$ to determine finite volume 
effects in the quark and gluon propagators 
\cite{Fischer:2002eq,Fischer:2005ui,Fischer:2007pf}}.

This work is organised as follows. In section \ref{sec2} we recall results
for QED$_3$ in the infinite volume/continuum limit. We present the 
Dyson-Schwinger equations for the fermion and photon propagators and discuss 
their asymptotic properties. In section \ref{sec3} we recall general
conditions for chiral symmetry breaking on a finite volume, formulate the DSEs
in a box and discuss our numerical methods to solve these. In section \ref{sec4}
we present our numerical results for the propagators at finite volume and the
critical number of flavours $N_f^c$ as a function of the box size. In
section \ref{sec5} we conclude.

\section{\label{sec2} $\mbox{QED}_3$ in the infinite volume/continuum limit}

\subsection{The Dyson--Schwinger equations in $\mbox{QED}_3$}

We consider $\mbox{QED}_3$ with a four-component spinor 
representation for the Dirac algebra and $N_f$ fermions.  
This allows a definition of chiral symmetry similar to 
the cases of $\mbox{QED}_4$ and $\mbox{QCD}_4$.  With 
massless fermions, the Lagrangian has a $U(2N_f)$ 
``chiral'' symmetry, which is broken to $SU(N_f) \times
SU(N_f) \times U(1) \times U(1)$ if the fermions become
massive. The order parameter for this symmetry
breaking is the chiral condensate which can be determined
\emph{e.g.} via the fermion propagator.

\begin{figure}[t]
\centerline{ \epsfig{file=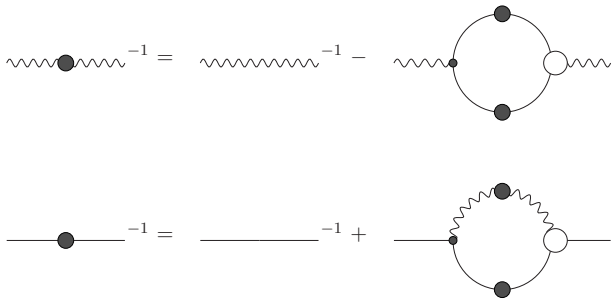,width=\columnwidth} }
\caption{\label{dse}
The Dyson--Schwinger equations of the photon and fermion 
propagators in diagrammatic notation. Wiggly lines denote photon
propagators, straight lines the fermion. A black dot denotes a 
bare fermion-photon vertex while the open circle denotes a 
dressed one.}
\end{figure}

The Dyson-Schwinger equations for the photon and fermion 
propagators in Euclidean space are given diagrammatically 
in Fig.~\ref{dse}. 
They read explicitly
\begin{eqnarray}
D^{-1}_{\mu \nu}(p) &=& D^{-1}_{0,\mu \nu}(p) \label{fermionDSE}\\
 &-&
 Z_1 N_f e^2 \int \frac{d^3q}{(2 \pi)^3} 
 \mbox{Tr}\left[ \gamma_\mu S(q) \Gamma_\nu(q,k) S(k) \right],
 \nonumber\\
S^{-1}(p) &=& S^{-1}_0(p) \nonumber\\
 &+&
 Z_1 e^2 \int \frac{d^3q}{(2\pi)^3}  \gamma_\mu  S(q) 
 \Gamma_\nu(q,p)\: D_{\mu \nu}(k), \label{photonDSE}
\end{eqnarray}
with the momentum routing $k_\mu = q_\mu -p_\mu$. Here
$N_f$ denotes the number of fermion flavours and $Z_1$ is the 
renormalisation constant of the fermion-photon vertex $\Gamma_\nu$.
  
In Landau gauge the general form of the dressed fermion propagator
$S(p)$ and the photon propagator $D_{\mu \nu}(p)$ is given by
\begin{eqnarray}
 S(p) &=& \frac{i \pslash A(p^2) + B(p^2)}
                   {p^2 A^2(p^2) + B^2(p^2)} \,,
\\
 D_{\mu \nu}(p) &=&
 \left(\delta_{\mu \nu} - \frac{p_\mu p_\nu}{p^2} \right) \,
 \frac{G(p^2)}{p^2} \, ,\nonumber\\
  &=& 
 \left(\delta_{\mu \nu} - \frac{p_\mu p_\nu}{p^2} \right) \,
 \frac{1}{p^2 (Z_3+\Pi(p^2))} \, ,
\label{photon}
\end{eqnarray} 
with the photon dressing function $G(p^2)$, the photon polarisation 
$\Pi(p^2)$ and renormalisation function $Z_3$ and the fermion dressing 
functions $A(p^2)$ and $B(p^2)$. These can be 
rearranged into the renormalisation group invariant fermion mass 
function $M(p^2)=B(p^2)/A(p^2)$ and the fermion wave function 
$Z_f(p^2)=1/A(p^2)$. Another renormalisation group invariant is 
the `running coupling' $e^2 \, G(p^2)$ built from the renormalisation 
point dependent photon dressing function $G(p^2)$ and the renormalised 
coupling $e^2$. The bare renormalised fermion propagator is given by 
$S^{-1}_0(p) = Z_2 (i \pslash + Z_m m)$ where $m$ is the renormalised 
current fermion mass, $Z_m$ is the mass renormalisation function and $Z_2$
the corresponding one for the fermion wave function. Note that in QED we
have the Ward-Takahashi identity $Z_1 = Z_2$.

The gauge dependence of the fermion mass and wave function and,
correspondingly, the fermion-photon vertex has been a much debated 
issue in the past, see \emph{e.g.} 
\cite{Fischer:2004nq,Curtis:1990zs,Burden:gy,Dong:jr,Kizilersu:1995iz,Bashir:2001vi,
Bashir:2002dz,Bashir:2002sp,Bashir:2004yt,Bashir:2008fk} 
and references therein. There in particular the technical question
of how to truncate the fermion-photon interaction to obtain
a gauge covariant pattern of chiral symmetry breaking and restoration
has been discussed. This issue is not quite settled yet, however
there are clear indications \cite{Fischer:2004nq,Bashir:2008fk,nedpaw} that 
Landau gauge is preferred in the sense that it allows for particularly
simple approximation schemes\footnote{In all other linear covariant 
gauges parametrised by the gauge parameter $\xi$ a gauge dependent 
scale $\frac{\xi e^2}{8\pi}$ appears which complicates matters 
considerably.}. This is why we prefer Landau gauge in this work.

A range of ans\"atze for the fermion-photon vertex have been 
investigated in \cite{Fischer:2004nq}. There it has been 
found that the critical number of flavours $N_f^c$ obtained with
the most elaborate construction, obeying the Ward-Takahashi identity, 
is almost similar to the one of the most
simple ansatz, a bare renormalised vertex 
\begin{equation}
\Gamma_\nu = Z_1 \gamma_\nu\,. \label{vertex}
\end{equation}
Therefore to keep matters as simple as possible we will only 
present results for the bare vertex approximation in this work. 
We did check, however, that more sophisticated ans\"atze do not 
alter the main conclusions presented below. In addition, a general
analysis of the infrared behaviour of QED$_3$ in the framework of
functional renormalisation group equations indicates that in
Landau gauge the bare vertex (\ref{vertex}) may even be the best 
possible choice \cite{nedpaw}.

Substituting this vertex into the fermion and photon DSEs,
taking appropriate traces and contracting the photon DSE with
the projector
\begin{equation}
{\mathcal P}_{\mu\nu} (p) = \delta_{\mu\nu} - 
\zeta\frac{p_\mu p_\nu}{p^2} 
\end{equation}
with $\zeta=1$ we arrive at
\begin{widetext}
  \begin{eqnarray}
B(p^2) \; &=& \; Z_2 Z_m m + Z_2^2  e^2 \int \frac{d^3q}{(2\pi)^3} 
  \frac{2 B(q^2)}{q^2 A^2(q^2) + B^2(q^2)} 
  \frac{G(k^2)}{k^2}   
\label{B-eq}\\  
A(p^2) \;&=&\; Z_2 +  Z_2^2 e^2  \int \frac{d^3q}{(2\pi)^3} 
  \frac{A(q^2)}{q^2 A^2(q^2) + B^2(q^2)}  
 \frac{G(k^2)}{k^2}  
 \left(  -\frac{k^2}{2p^2} + \frac{(p^2-q^2)^2}{2k^2p^2} \right)
\label{A-eq}\\
\frac{1}{G(p^2)} = Z_3 + \Pi(p^2) &=& Z_3 - Z_2^2 e^2 N_f \int \frac{d^3q}{(2 \pi)^3} 
\; \frac{1}{q^2 A^2(q^2) + B^2(q^2)} \; \frac{1}{k^2 A^2(k^2) + B^2(k^2)} 
\widetilde{W}_1(p^2,q^2,k^2) 
\label{Pi-eq}
\end{eqnarray}
\end{widetext}
where we used $Z_1=Z_2$ and the subtracted kernel 
\begin{equation}
  \widetilde{W}_1\left( p^2,q^2,k^2 \right)=
  W_1\left(p^2,q^2,k^2\right)-\frac{2k^2\left( 3-\zeta \right)}{3p^2} \,,
\label{lindiv}
\end{equation}
with
\begin{eqnarray}
W_1\left(p^2,q^2,k^2\right) &=&
\frac{\zeta k^4}{ p^4}+k^2\left(
\frac{1-\zeta}{p^2}-\frac{2\zeta q^2}{p^4}
\right)-1\nonumber\\[-2mm] \label{w1}\\[-2mm]
&+&\frac{\left( 1-\zeta \right)q^2}{p^2}+\frac{\zeta q^4}{p^4}
\nonumber
\end{eqnarray}
in the photon equation. As explained in appendix A of ref.~\cite{Fischer:2004nq}
the subtraction of the term proportional to $(3-\zeta)$ in Eq.~(\ref{lindiv})
is necessary to avoid spurious linear divergences in the photon-DSE generated
by the regularisation procedure (a hard cutoff) used in the numerical treatment of the equations.
The choice $\zeta=1$, \emph{i.e.} the transverse projection of the photon equation, is
mandatory to avoid the back-reaction of spurious longitudinal terms into the
right hand side of the photon equation. Compared to \cite{Fischer:2004nq}, where
$\zeta=3$ has been used, this treatment leads to a quantitatively improved value 
for the critical number of flavours, see below. Note, however, that the effects
of varying $\zeta$ are quantitatively small in general; all qualitative conclusions 
derived in \cite{Fischer:2004nq} and also here are independent of the choice of $\zeta$.

\subsection{Asymptotic behaviour of the propagators}

An often used approximation to determine the 
asymptotic behaviour of the fermion and photon dressing 
functions has been the $1/N_f$-expansion. This expansion
is equivalent to a perturbative expansion for small $e^2$ 
while keeping $\alpha = N_f e^2/8$ fixed. As $\mbox{QED}_3$ 
is an asymptotically free theory this expansion does
provide correct answers in the ultraviolet. For the photon
polarisation and the vector dressing function of the fermion
one finds for $N_f$ massless fermion
flavours \cite{Appelquist:1986fd}
\begin{eqnarray}
\Pi(p^2 \gg \alpha) &=& \frac{N_f \; e^2}{8p} 
= \frac{\alpha}{p} \, , \label{UV_PH} \\
A(p^2 \gg \alpha) &=& 1 \label{UV_A}
\end{eqnarray}
For the mass function of the fermion one can
use the operator product expansion to obtain 
\begin{equation}
 M(p^2 \gg \alpha) = 
 \frac{2+\xi}{4} \; \frac{\langle\bar{\Psi} \Psi \rangle}{p^2}  \,,
\label{UV_B}
\end{equation}
with the chiral condensate $\langle\bar{\Psi} \Psi \rangle$. Note
that the condensate can also be determined from the trace of
the fermion propagator:
\begin{equation}
 \langle \bar{\Psi} \Psi \rangle =  
  -4 Z_2 \int \frac{d^3q}{(2 \pi)^3} 
 \frac{B(q^2)}{q^2 A^2(q^2) + B^2(q^2)} \,.
\label{trace-condensate}
\end{equation}

In the infrared momentum region the $1/N_f$-expansion is 
clearly not sufficient. Here one has to resort to a self-consistent
analysis of the DSEs in terms of asymptotic power-laws. This method 
is well developed in QCD$_4$ 
\cite{Lerche:2002ep,Zwanziger:2002ia,Alkofer:2004it,Fischer:2006vf} 
and has been adapted to $\mbox{QED}_3$ in Ref.~\cite{Fischer:2004nq}. 
Such an analysis is valid if no scales are present, \emph{i.e.} in 
the deep infrared momentum region $p \ll \alpha$ and only in the absence 
of fermion masses. The appearance of infrared power laws with potentially 
fractional anomalous dimensions is thus an indicative and characteristic 
property of the symmetric phase of QED$_3$. 

For a bare fermion-photon vertex the infrared behaviour of the vector 
fermion dressing function and the photon polarisation in the chirally 
symmetric phase can be written as
\begin{eqnarray}
A(p^2) \sim p^{2\kappa} \,, \hspace*{5mm} \Pi(p^2) \sim p^{-1-4\kappa}\,.
\end{eqnarray}
In case of a dressed fermion-photon vertex according to the
Ward-Takahashi identity the corresponding power-laws are
\begin{eqnarray}
A(p^2) \sim p^{2\kappa} \,, \hspace*{5mm} \Pi(p^2) \sim p^{-1-2\kappa}\,.
\end{eqnarray}
These expressions solve the DSEs in the chirally symmetric phase
as described in detail in \cite{Fischer:2004nq} (see also
\cite{Bashir:2008fk} for a re-derivation from a slightly different
perspective).

\begin{figure}[t]
\epsfig{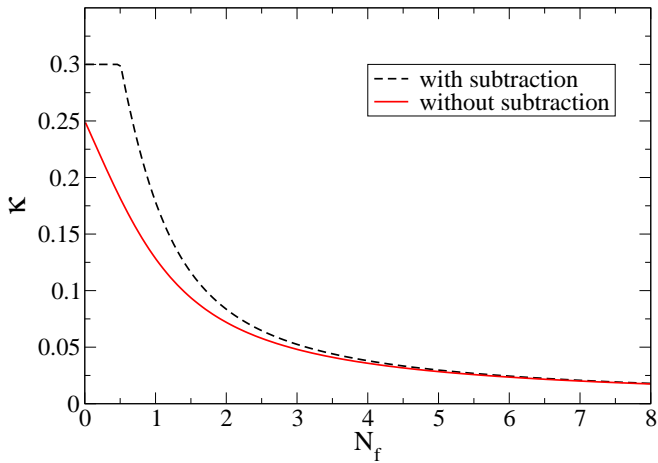}\vspace*{4mm} 
\caption{The infrared exponent $\kappa$ as a function of the 
number of flavours $N_f$ in the symmetric phase of QED$_3$. }
\label{fig:kappa}
\end{figure}

The value of $\kappa$ as a function of $N_f$ can then be determined
from the DSEs using a procedure given in detail in \cite{Fischer:2004nq}. 
The solutions for a bare vertex with and without the subtracted term in
Eq.~(\ref{lindiv}) is plotted in Fig.~\ref{fig:kappa}. For $N_f > 1.5$ the
exact solutions can be represented by the fits 
\begin{eqnarray} 
 \kappa_{bare}^{sub} &=& \frac{0.142}{N_f} + \frac{0.002}{N_f^2} +  O(1/N_f^3) \, ,\label{kappa_sub}\\
 \kappa_{bare}       &=& \frac{0.137}{N_f} + \frac{0.058}{N_f^2} +  O(1/N_f^3) \, .\label{kappa_wosub}
\label{kappa_bare}
\end{eqnarray}
A corresponding fit for the bare vertex and a Ward-Identity improved
vertex for $\zeta=3$ is given in \cite{Fischer:2004nq}. The difference
between (\ref{kappa_sub}) and (\ref{kappa_wosub}) represents the systematic
uncertainty in our numerical calculation due to the cutoff regularisation
procedure used. For dimensional regularisation the subtraction of the term
in (\ref{lindiv}) is not necessary and therefore the unsubtracted result
Eq.~(\ref{kappa_wosub}) should be viewed as the exact one for the bare
vertex truncation scheme. In principle one can reproduce
(\ref{kappa_wosub}) also in a numerical treatment of the DSEs with a hard 
cutoff when vertex corrections around the cutoff scale are included, 
see \cite{Fischer:2008uz} for details. However, this procedure is much more 
involved than the simple subtraction scheme of Eq.~(\ref{lindiv}). 
Since the difference between (\ref{kappa_sub}) and (\ref{kappa_wosub}) is 
rather small in the interesting region above $N_f=2$ and therefore unimportant 
for all of the conclusions of the present work we resort to the simple 
subtraction scheme (\ref{lindiv}) and consequently reproduce (\ref{kappa_sub}) 
in our numerics.

The critical number of flavours $N_f^c$ where chiral symmetry
is restored can be determined analytically from the DSE for the
scalar fermion dressing function $B(p^2)$. Again we refer the reader
for details to Ref.~\cite{Fischer:2004nq} and merely state the result
\begin{eqnarray}
(N_f^c)_{bare}^{sub} &\approx& 3.84 \, , \label{crit}\\
(N_f^c)_{bare}       &\approx& 3.61 \, .
\end{eqnarray}
The corresponding result from the functional renormalisation group is
$(N_f^c)_{bare} \approx 3.6$ \cite{nedpaw} in agreement with our 
result for the unsubtracted equation.
Numerical results for Ward-identity improved vertices lead to 
results in the range of $3.5-4$ \cite{Fischer:2004nq}. These numbers 
can be contrasted with $(N_f^c)_{1/N_f} = 32/\pi^2 \approx 3.24$ 
from the $1/N_f$-expansion \cite{Appelquist:1988sr}. Note that all these
results are far above $N_f=2$ relevant for the description of 
high-$T_c$ cuprate superconductors as discussed in the introduction.

Finally we wish to emphasise that these power law solutions can only
be obtained when the full structure of the propagator DSEs is taken
into account. The authors of 
Refs.~\cite{Atkinson:1989fp,Pennington:1990bx}
did not find a solution corresponding to a symmetric phase in their
truncation scheme because the feedback from the function $A$ onto 
the vacuum polarisation is not taken into account. This then prohibits
the appearance of power laws and therefore does not allow for the
appearance of the chirally symmetric phase and should be 
discarded \cite{Fischer:2004nq,Bashir:2008fk}. 

\section{\label{sec3} $\mbox{QED}_3$ at finite volumes}

\subsection{Chiral symmetry breaking in a box \label{sec:chiral}}

Before we embark on our investigation, let us recall the general finite 
volume behaviour of the chiral condensate \cite{Leutwyler:1992yt}.
The fermion propagator in its spectral representation is given by
\begin{equation}
S_A(x,y) \;=\; \sum_n \frac{u_n(x)\, u^\dagger_n(y)}{m-i\lambda},
\end{equation}
where $u_n(x)$ and $\lambda_n$ are eigenfunctions and eigenvalues of the
Euclidean Dirac operator, $\Dslash u_n(x) = \lambda_n u_n(x)$. The gauge field
$A$ is treated as an external field. These 
eigenfunctions occur either as zero modes or in pairs of opposite eigenvalues.
Setting $x=y$, integrating over x and neglecting the zero mode contributions,
one obtains
\begin{equation}
\frac{1}{V} \int_V S_A(x,x) \;=\; -\frac{2m}{V} \sum_{\lambda_n > 0} 
\frac{1}{m^2+\lambda_n^2}. \label{eq2}
\end{equation}
The chiral condensate can be deduced by averaging the left hand side of this 
equation over all gauge field configurations and then taking the infinite 
volume limit to give
\begin{equation}
\langle \bar{q}q\rangle \;=\; -2m \int^\infty_0 d\lambda \;
\frac{\rho(\lambda)}{m^2 + \lambda^2}, 
\end{equation}
where $\rho(\lambda)$ is the mean level density of the spectrum, which becomes 
dense in the infinite volume limit. In the chiral limit, $m \rightarrow 0$, only
the infrared part of the spectrum contributes and one finally arrives at the 
Banks-Casher relation \cite{Banks:1979yr}
\begin{equation}
\langle \bar{q}q\rangle \;=\; -\pi \rho(0) \label{Banks-Casher}\; .
\end{equation}

If the two limits are interchanged, \emph{i.e.} if one takes 
the chiral limit before the infinite volume limit, one has a discrete 
sum in Eq.~(\ref{eq2}) and the infrared part of the spectrum cannot 
trigger a non-vanishing chiral condensate: chiral symmetry is restored. 
If, however, at a given volume the explicit fermion mass $m$ is not too small, 
one can still observe the spontaneous formation of a quark condensate. 
If the factor $(m^2+\lambda_n^2)^{-1}$ varies only slightly with $n$,
the sum in Eq.~(\ref{eq2}) can still be replaced by an integral and 
Eq.~(\ref{Banks-Casher}) remains valid. For this to be a legitimate 
approximation one needs 
$m \gg \Delta \lambda \sim 1/V\rho(\lambda) = \pi/(V|\langle \bar{q}q\rangle|)$,
at the lower end of the spectrum. Thus one obtains the condition \cite{Leutwyler:1992yt}
\begin{equation}
V m |\langle \bar{q}q\rangle| \gg \pi. \label{torusrelation}
\end{equation}
This relation reveals the crux of the matter. In principle, if the volume is 
large enough, the necessary quark masses are academically small and may even be 
neglected in the numerical treatment. What counts as {\it large} in this context,
however, depends sensitively on the size of the chiral condensate. In QED$_3$ the
condensate becomes extremely small already well below the critical number of
flavours $N_f^c$ of the chiral phase transition \cite{Fischer:2004nq}. 
Thus although a formulation on a finite volume may do well for $N_f = 1$,
all signals of dynamical chiral symmetry breaking will be lost already well below
the $N_f^c$ of the theory in the infinite volume/continuum limit. This
behaviour will be quantified below.

\subsection{The DSEs on a torus \label{sec:dsetorus}}

On a compact manifold, the photon and fermion fields have to obey appropriate
boundary conditions in the time direction. These have to be periodic for
the photon fields and antiperiodic for the fermions. For computational reasons 
it is highly advantageous, though not necessary, to choose the same conditions 
in the spatial directions. We 
choose the box to be of equal length in all directions, $L_1=L_2=L_3\equiv L$, 
and denote the corresponding volume $V=L^3$. Together with the boundary conditions 
this leads to discretised momenta in momentum space. Thus all momentum integrals 
appearing in the Dyson-Schwinger equations are replaced by sums over Matsubara modes. 

\begin{figure}[t]
\centerline{\epsfig{file=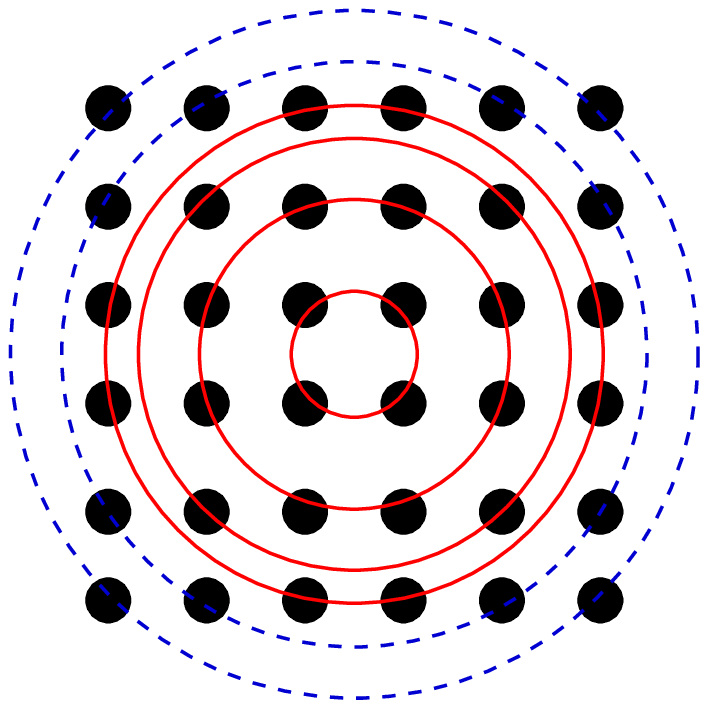,width=6cm}}
\caption{Two-dimensional sketch of the momentum grid dual to the three-torus
for a fixed Cartesian momentum cutoff. The hyperspheres depicted by dashed 
lines are not complete in the sense that additional momentum points on
these spheres are generated if the cutoff is increased. The torus equivalent 
of an O(3)-invariant cutoff used in our calculations sums only over complete 
hyperspheres, which are indicated by fully drawn circles.}
\label{fig:latt}
\end{figure}

On a torus with antiperiodic boundary conditions for the fermion fields, the
momentum integral changes into a sum of Matsubara modes,
\begin{equation}
\int \frac{d^3q}{(2 \pi)^3} \:(\cdots) \:\:  \longrightarrow \:\:\frac{1}{L^3}
\sum_{n_1,n_2,n_3} \:(\cdots) \,, 
\end{equation}
counting momenta ${\bf q}_{\bf n} = \sum_{i=1..3} (2\pi/L)(n_i+1/2) \hat{e}_i$, where
$\hat{e}_i$ are Cartesian unit vectors in Euclidean momentum space. For the
photon with periodic boundary conditions the momentum counting goes like 
${\bf q}_{\bf n} = \sum_{i=1..3} (2\pi/L)(n_i) \hat{e}_i$. For the numerical
treatment of the equations it is convenient to rearrange these summations 
such that they represent a spherical coordinate system \cite{Fischer:2002eq}, 
see Fig.~\ref{fig:latt} for an illustration. We then write
\vspace{2mm}  
\begin{equation}
\frac{1}{L^3} \sum_{n_1,n_2,n_3} (\cdots) \: =  \frac{1}{L^3} \sum_{j,m} \:(\cdots) \,, 
\end{equation}
where $j$ counts spheres with ${\bf q_n q_n}=\textrm{const}$, and 
$m$ numbers the grid points on a given sphere. The corresponding momentum vectors are
denoted ${\bf q}_{m,j}$ and their absolute values are given by 
$q_{m,j} = |{\bf q}_{m,j}|$. It is then a simple matter to
introduce the torus equivalent of an $O(3)$-invariant cut-off by 
restricting $j$ to an interval $\left[1,\mathrm{N}\right]$. This procedure
is equivalent to `cutting the edges' of our torus as indicated in Fig.~\ref{fig:latt}.

The resulting DSEs are then given by 
\begin{widetext}
  \begin{eqnarray}
B(p_{i,l}^2) \; &=& \; Z_2 Z_m m + Z_2^2 \: \frac{e^2}{L^3} \sum_{j,m}^N 
  \frac{2 B(q_{j,m}^2)}{q_{j,m}^2 A^2(q_{j,m}^2) + B^2(q_{j,m}^2)} 
  \frac{G(k^2_{i,l,j,m})}{k_{i,l,j,m}^2}   
\label{B-eq-t}\\ [2mm]
A(p_{i,l}^2) \;&=&\; Z_2 +  Z_2^2 \:\frac{e^2}{L^3} \sum_{j,m}^N
  \frac{A(q_{j,m}^2)}{q_{j,m}^2 A^2(q_{j,m}^2) + B^2(q_{j,m}^2)}  
 \frac{G(k^2_{i,l,j,m})}{k_{i,l,j,m}^2}  
 \left(  -\frac{k_{i,l,j,m}^2}{2p_{i,l}^2} 
 + \frac{(p_{i,l}^2-q_{j,m}^2)^2}{2k_{i,l,j,m}^2p_{i,l}^2} \right)
\label{A-eq-t}\\[2mm]
 \frac{1}{G(p_{i,l}^2)} 
 &=&Z_3-Z_2^2 \: \frac{e^2 N_f}{L^3} \sum_{j,m}^N
\; \frac{1}{q_{j,m}^2 A^2(q_{j,m}^2) + B^2(q_{j,m}^2)} \; \frac{1}{k_{i,l,j,m}^2 A^2(k_{i,l,j,m}^2) + B^2(k_{i,l,j,m}^2)} 
\widetilde{W}_1(p_{i,l}^2,q_{j,m}^2,k_{i,l,j,m}^2)\,. 
\label{Pi-eq-t}
\end{eqnarray}
\end{widetext}
Note that the momentum argument $k$ of the photon self-energy in the fermion DSE
is the difference $k_{i,l,j,q} = p_{i,l}-q_{j,q}$ of two antiperiodic 
Matsubara momenta and thus lives on a momentum grid corresponding to periodic 
boundary conditions, as it should.

The DSEs can be solved numerically employing well established methods. 
Our numerical method on the torus is outlined in Ref.~\cite{Fischer:2005ui}, 
the corresponding continuum method as well as details on the renormalisation 
procedure of the DSEs are given in Ref.~\cite{Fischer:2004nq}. 

Note that the propagators determined from the continuum version of the DSEs, 
Eqs.~(\ref{B-eq}--\ref{Pi-eq})
are independent of the regularisation procedure. In our numerical calculations 
in the infinite volume/continuum limit we 
use a subtracted version of these equations and an O(3)-invariant UV-cutoff 
which can be sent to infinity at the end of each calculation. These DSEs therefore 
represent not only the infinite volume limit but also the continuum limit 
(in coordinate space) of the representation given by Eqs.~(\ref{B-eq}--\ref{Pi-eq}) of 
the DSEs on a torus. We use the phrase infinite volume/continuum 
limit to indicate this simultaneous removal of both an ultraviolet and an infrared 
cutoff.

\section{Numerical results \label{sec4}}

\subsection{Finite size effects: fixing the UV cutoff \label{res:UV}}

\begin{figure}[t]
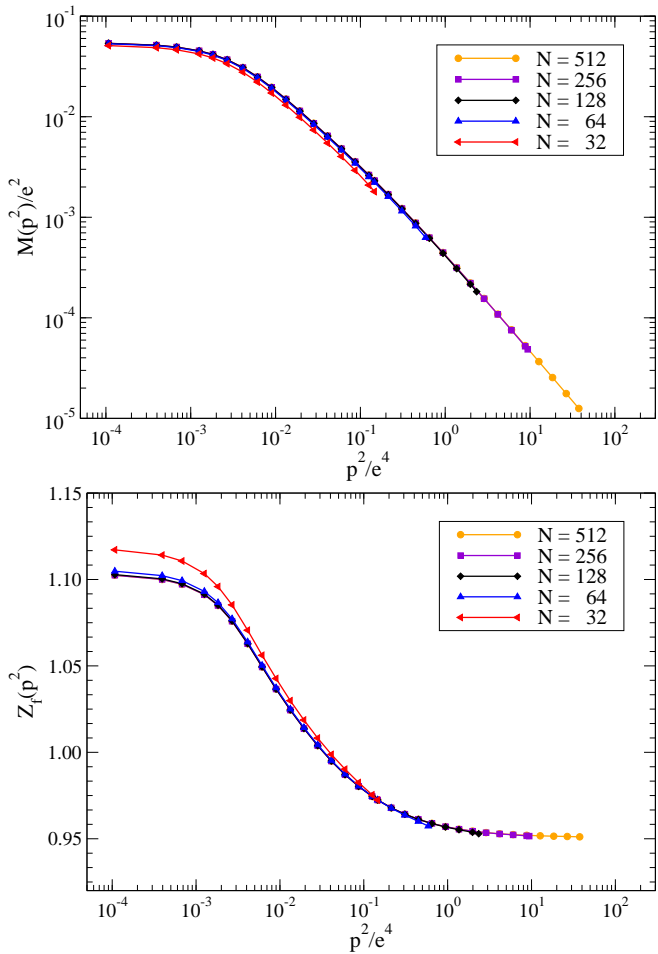

\epsfig{file=plot_IR_fixed_UV_vary_M.eps,width=\columnwidth} 
\epsfig{file=plot_IR_fixed_UV_vary_Z.eps,width=\columnwidth}
\caption{Fermion mass function $M(p^2)$ and wave function $Z_f(p^2)$
as a function of scaled momentum $p^2/e^4$. The volume of the box
corresponding to the smallest available momentum is kept fixed and the
ultraviolet cutoff is varied. Here $N$ counts the number of 
momentum points in one direction of our torus, \emph{i.e.} $N=100$ means 
that we employed a torus with $200^3$ momentum vectors. This
translates into the cutoffs 
$\Lambda_{UV}^2 = 0.15 \,e^4, 0.62 \,e^4, 2.47 \,e^4, 9.86 \,e^4, 39.5
\,e^4$.} 
\label{fig:UV}
\end{figure}

Before we investigate the finite volume effects of various quantities of
interest we have to clarify whether there are sizeable effects due to the
finite size of the system corresponding to a fixed ultraviolet momentum 
cutoff. Part of these effects are removed by our choice of
`cutting the edges' of the torus, described in the previous section. The
remaining finite size effects can be evaluated by varying the size of the
cutoff. Here, a natural minimal cutoff is given by the intrinsic scale
$\alpha = N_f \, e^2/8$ of QED$_3$. Thus, working in the range $0< N_f \le 8$, 
we anticipate that finite size effects are of minor importance for 
cutoffs larger than the intrinsic scale, \emph{i.e.} $\Lambda^2 \ge e^4$. Indeed, 
this is the case
as can be seen from the two plots in Fig.~\ref{fig:UV}. Shown are the
fermion mass function $M(p^2)$ and wave function $Z_f(p^2)$ as a function 
of scaled momentum $p^2/e^4$ for various cutoffs between 
$\Lambda_{UV}^2=0.15 \,e^4$ and $\Lambda_{UV}^2=39.5 \,e^4$. 
The number of fermion
flavours is chosen to be $N_f=1$, similar results are obtained for other choices.
The box volume for the fermions is related to the lowest momentum point $p^2_{min}$
by $V=L^3$ with $L = \sqrt{3} \pi/p_{min}$. Here we use $L = 390/e^2$.
Note that for the photon the box volume is related to the lowest momentum point by 
$V=L^3$ with $L = 2 \pi/p_{min}$. The same box length then results
in a slightly different value for the lowest momentum point in the 
photon dressing function. We refrain from showing the photon explicitly
here, since the finite size effects are similar to the ones for the fermions.

In Fig.~\ref{fig:UV} the finite size effects are visible only for the smallest 
cutoff. There are effects in both the infrared and ultraviolet momentum regions,
where a number of momentum points deviate from the results with larger cutoffs. 
This is true for the renormalisation point dependent fermion wave function, 
normalised such that they match the continuum results, 
but also for 
the renormalisation point independent fermion mass function. The perhaps 
surprising observation that the variation of an ultraviolet cutoff also affects 
the infrared behaviour of the dressing functions implies a certain entanglement 
between the infrared and ultraviolet modes of a gauge theory. Similar effects have 
been found in four dimensional Yang-Mills theory, see \emph{e.g.} \cite{Fischer:2008uz} 
and references therein.

We also need to comment on the fact that we observe dynamical chiral 
symmetry breaking in our system at all, despite in practice working with a 
vanishing bare quark mass. In the continuum formulation this entails 
working in the chiral limit. However, due to the formal reasons discussed in section
\ref{sec:chiral} this cannot be true on a finite volume, since taking the
chiral limit before the infinite volume limit inevitably leads to the loss 
of dynamical chiral symmetry breaking. However, this is not what we observe here.
Indeed, the volumes we use are large enough to allow for extremely small 
bare fermion masses according to the relation of Eq.~(\ref{torusrelation}). 
In practice, this allows one to neglect the fermion mass entirely in the numerical 
treatment of the DSEs. However it is important to keep in mind that this 
means we are working \emph{close} to the chiral limit, but not \emph{in} the 
chiral limit. 

We conclude this subsection with the observation that from about 
$\Lambda_{UV}^2 = e^4$ onwards almost no finite size effects are present.
We therefore consider this value a lower bound for admissible cutoffs on 
a torus, in agreement with our general considerations concerning the natural 
scale in QED$_3$. This finding also agrees with corresponding results on
finite size effects in lattice simulations, see \cite{Strouthos:2008hs} 
and Refs. therein. In order to be absolutely sure that cutoff effects do not
play any role in what follows we use the somewhat larger cutoff 
$\Lambda_{UV}^2 = 2.35 \,e^4$ from now on. 

\subsection{Finite volume effects: towards the infinite volume limit}

\begin{figure}[t]
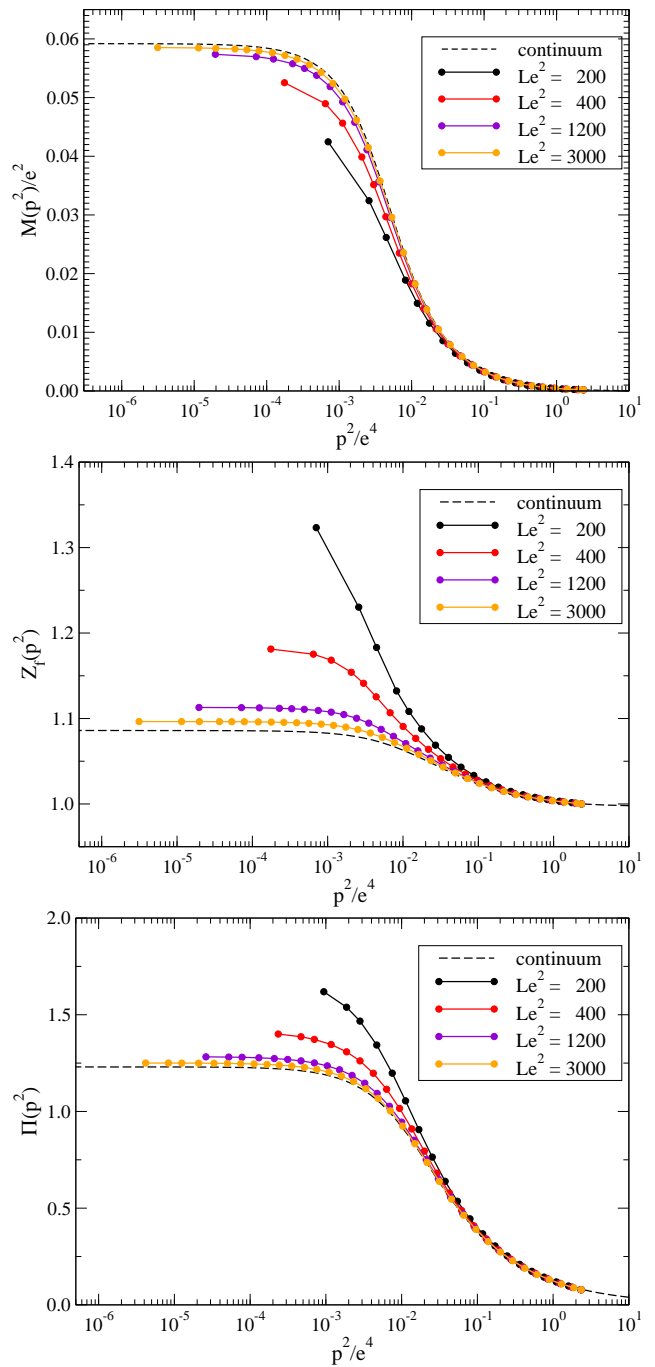

\epsfig{file=plot_IR_scaling_x4_M.eps,width=0.97\columnwidth} 
\epsfig{file=plot_IR_scaling_x4_Z.eps,width=0.97\columnwidth} 
\epsfig{file=plot_IR_scaling_x4_PI.eps,width=0.97\columnwidth} 
\caption{Fermion mass function $M(p^2)$, wave function $Z_f(p^2)$
and photon self energy $\Pi(p^2)$ 
as a function of scaled 
momentum $p^2/e^4$. The ultraviolet momentum cutoff is kept fixed
at $\Lambda_{UV}^2 = 2.35 \,e^4$ and the box length is varied from 
$Le^2=200-3000$.}
\label{fig:IR}
\end{figure}

We are now in a position to study the finite volume effects occurring for the
fermion and photon propagators on a torus. To this end we keep the ultraviolet 
cutoff of our system fixed and vary the infrared cutoff in the range of 
$L\,e^2=200-3000$.
We show the resulting behaviour of the fermion mass and wave function as well as
the photon together
with the corresponding continuum results  
in Fig.~\ref{fig:IR}. Again, we choose $N_f=1$. The variation
of the volume clearly results in the loss of a substantial amount of
generated fermion mass when the volume gets smaller and smaller. For even smaller 
volumes than shown in the figure chiral symmetry is restored in agreement with
the condition $V m |\langle \bar{q}q\rangle| \gg \pi$
discussed in the subsections \ref{sec:chiral} and \ref{res:UV}. 
On the other hand, we observe that extremely large volumes are needed to
account for the full effect of dynamical chiral symmetry breaking observed in
the infinite volume/continuum limit. This is in marked contrast with the
behaviour of the quark sector of QCD$_4$ \cite{Fischer:2005nf} and can be explained by
a closer look at the scales involved. Whereas in QCD$_4$ the generated quark masses
($M(0) \approx 300-400 \,\mbox{MeV}$) are of the same order as the intrinsic scale 
of the system ($\Lambda_{QCD}^{\bar{MS}} \approx 250 \,\mbox{MeV}$), the situation
is clearly different in QED$_3$. Here our characteristic scale is of the order 
$\alpha = e^2/8$, whereas the generated fermion masses are of order $10^{-2} \,e^2$, 
as can be seen from the plot. To keep volume effects small, this scale has to be well 
accommodated by the system on a box, which translates to a lowest momentum to be 
much smaller than $p^2 = 10^{-5} \,e^4$. Indeed, this is what we observe. Choosing
the volume large enough that the lowest momentum is well below this scale we approach
the infinite volume/continuum limit. This is true for all three dressing functions, 
the fermion mass function $M(p^2)$, the fermion wave function $Z_f(p^2)$ and the 
photon self energy $\Pi(p^2)$.

\subsection{Finite volume effects: $Z_f(N_f)$ and $\kappa(N_f)$}

Let us now examine the 
influence of finite volume effects on the critical value $N_f^c$,
where the system undergoes a phase transition from the chirally broken
into the chirally symmetric phase. This phase transition is marked
by the change of the infrared asymptotics of the fermion wave 
function $Z_f(p^2)$; for $N_f < N_f^c$ this function is a constant
in the infrared, whereas for $N_f > N_f^c$ it develops a power law
with $N_f$-dependent exponent $\kappa$ \cite{Fischer:2004nq}. 
In a sense, $\kappa$ can be viewed as an order parameter of this
phase transition\footnote{Note, however, that the phase transition is
not a first or second order transition but has similar properties
to the conformal transition of QED in four dimensions 
\cite{Fischer:2004nq,Miransky:1996pd}.}. We exhibit this behaviour
of $\kappa$ by fitting a power law in the infrared to the fermion
wave function 
\begin{align}
 Z_f(p^2)& = C \, (p^2)^{-\kappa}. \label{powerlaw}
\end{align}
which is related to the fermion vector dressing function by $Z_f=1/A$.
Both, the coefficient $C$ and the power $\kappa$ may depend on $N_f$.

\begin{figure}[t]
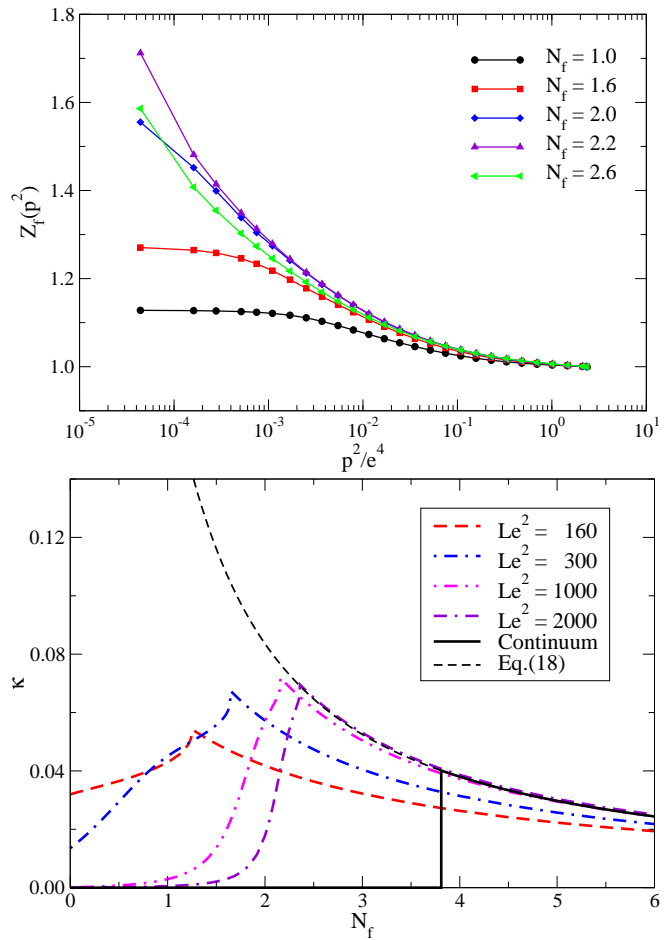

\epsfig{file=Zf_varNf.eps,width=\columnwidth} 
\epsfig{file=kappa.eps,width=\columnwidth} 
\caption{Upper plot: Fermion wave function $Z_f(p^2)$ as a function of 
scaled momentum $p^2/e^4$ and the number $N_f$ of fermion flavours. Note 
that for clarity of the figure from $p^2/e^4=10^{-3}$ upwards only results
for a selected number of momentum points on the torus are shown.
Lower plot: The resulting value of the infrared exponent $\kappa$
as a function of $N_f$ for different volumes.}
\label{fig:Zf}
\end{figure}

To visualize this procedure the fermion wave function $Z$ is 
shown in the upper panel of Fig.~\ref{fig:Zf} for a box with 
$L\,e^2 \approx 800$ for different 
numbers of flavours $N_f$. One clearly sees the aforementioned 
behaviour: for $N_f \le 2.0$ the function develops a constant 
in the infrared region (\emph{i.e.} $\kappa=0$), whose value is 
proportional to $N_f$. The phase transition occurs for this 
specific volume for $2.0 < N_f^c < 2.2$. For $N_f \ge 2.2$  one 
observes a power law of the wave function at small momenta. 
The coefficient of this power law is still proportional to $N_f$,
however the exponent $\kappa$ decreases with growing $N_f$ according
to Eq.~(\ref{kappa_sub}). Consequently we observe a decrease of the 
function $Z_f(p^2)$ with growing $N_f$ in the symmetric phase. As
a result one could determine $N_f^c$ as the number of flavours for
which $Z_f$ is maximal at a given infrared momentum. This is also
the case in the infinite volume/continuum limit \cite{Fischer:2004nq}. 

As an (equivalent) alternative we determine $N_f^c$ by fitting the
power law (\ref{powerlaw}) to our numerical results
for $Z_f(p^2)$ in the infrared momentum region. There is a caveat here: 
similar to four-dimensional Yang-Mills theory one observes that 
the power law (\ref{powerlaw}) can only be seen for momenta
$1/L \ll p \ll e^2$. This behaviour is generic on a torus \cite{Fischer:2007pf}.
To obtain significant results for our infrared coefficients we therefore 
have to perform the fit for momentum points significantly larger than 
the lowest one. In practice we chose momenta from the fourth point onwards. 

The resulting dependence of the exponent $\kappa$ on $N_f$ is plotted
in the lower panel of Fig.~\ref{fig:Zf} for several volumes of the box
together with the corresponding function in the infinite volume/continuum
limit. At infinite volume the function $\kappa(N_f)$ is zero for 
$N_f < N_f^c \approx 3.84$ and equal to the analytic result of 
Eq.~(\ref{crit}) in the symmetric phase above $N_f^c$. For a given finite
volume the phase transition is still indicated by a maximum in $\kappa(N_f)$. 
However, there is an additional region at $N_f < N_f^c(L)$, where the function 
rises slowly towards its maximal value. This region is generated by the finite
infrared cutoff of the system in a box, which prevents the fermion wave function
$Z_f(p^2)$ from bending towards a finite asymptotic value at $p^2=0$. This
effect mimics a power law at values of $N_f$ where the system is still in the
chirally broken phase. With increasing volume this region gets smaller and 
smaller until it reaches the sharp rise of the function in the infinite 
volume/continuum limit as shown in Fig.~\ref{fig:Zf}. 

As a result we find a critical number of flavours which depends
upon the volume of the torus. The explicit values are shown in 
Fig.~\ref{fig:massive} together with corresponding results from 
Ref.~\cite{Gusynin:2003ww}. In the following we concentrate on the solutions
close to the chiral limit and postpone the discussion of the (orange) curve 
with large bare fermion mass to the next subsection. 

For volumes that are not too large, our solutions and the ones reported
in Ref.~\cite{Gusynin:2003ww} can be well fitted by a form
\begin{equation}
N_f^c = a - b/(L\,e^2)^{1/3}\,. \label{fitt}
\end{equation}
For our results we obtain $a=3.23$ and $b=10.64$, shown as dashed line in the 
plot. Interestingly, to high precision the exponent of the box length in this
fit is given by $-1/3$, although we do not have a good explanation for exactly 
this behaviour. In terms of $(1/L)^{1/3}$ the fit then suggests a linear 
extrapolation to the infinite volume limit. However, it turns out that this 
linear behaviour breaks down for extremely large volumes. This is
evident for the results from Gusynin and Reenders \cite{Gusynin:2003ww}. 
The deviation from the linear behaviour is quantitatively
important: for the truncation of Gusynin and Reenders a linear extrapolation to the
infinite volume results in $N_f^c \approx 2.52$ as compared to the infinite volume
result $N_f^c = 3.2$. (Note that the curve gets extremely steep for the very largest
volumes, which are not plotted.) For our results we find a value of 
$N_f^c \approx 3.23$ for the linear extrapolation, whereas the analytical infinite 
volume result is $N_f^c \approx 3.84$. Though this difference is
not huge, it highlights the need for non-linear extrapolation procedures to the
infinite volume limit. Unfortunately our largest volumes are not yet large enough to
penetrate this region of nonlinearity.

\subsection{Non-vanishing bare fermion masses}

\begin{figure}[t]
\epsfig{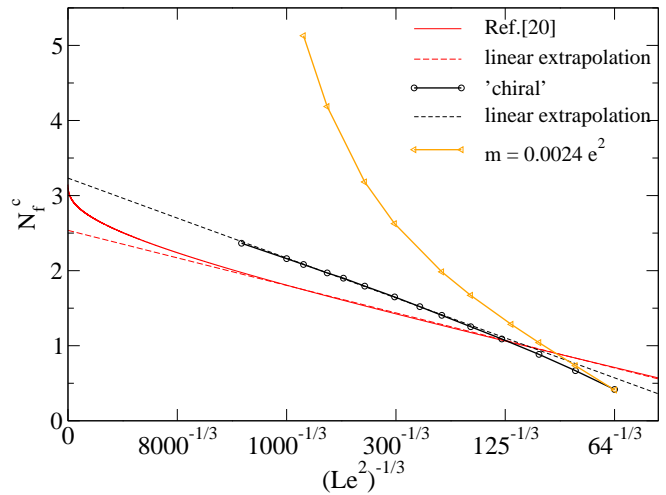} 
\caption{
The critical number of flavours $N_f^c$ as a 
function of inverse box length $1/L^{1/3}$. In the 'chiral' limit
we show the results of our calculation compared to the corresponding
one of Ref.~\cite{Gusynin:2003ww} in a simpler truncation scheme. The
dashed lines are linear extrapolations to the infinite volume limit. 
We also show results for a bare fermion mass comparable to the ones 
used in lattice calculations. }
\label{fig:massive}
\end{figure}

In this subsection we investigate the consequences of explicit bare fermion masses 
$m_0$ in the Lagrangian of QED$_3$ on the chiral phase transition. To this end
we first determined the properties of the system in the infinite volume/continuum 
limit by solving the corresponding DSEs for a range of different bare fermion  
masses. For any given mass and varying $N_f$ we find that the fermion wave 
function never develops a pure power law in the infrared. Instead there is a 
region $0 < p < \Lambda_{IR}$, where the function is constant and a region 
$\Lambda_{IR} < p < \alpha$ where a power law is present. The scale $\Lambda_{IR}$ 
depends on the explicit fermion mass as well as on the number of flavours. 
We conclude from this that away from the chiral limit the infrared exponent 
$\kappa$ is no longer an order parameter. As concerns the global behaviour of 
$Z_f(p^2)$ we still observe the behaviour discussed above Eq.~(\ref{fitt}):
at a small enough momentum $p^2$ the value of the function $Z_f(p^2)$ increases 
with $N_f$ up to a certain point at a critical value $N_f^c$ and decreases 
again if $N_f$ grows further. In the chiral limit this critical value $N_f^c$
marked the chiral phase transition. Here, however, this seems not to be so. 
The fermion mass function $M(0)$ decreases exponentially with $N_f$ without
a trace of a rapid change around $N_f^c$. We therefore conclude that there
is no phase transition in the infinite volume/continuum limit for QED$_3$ with
non-vanishing bare fermion masses $m_0$.

On a torus with a given volume, however, the scale $\Lambda_{IR}(N_f)$ 
decreases with $N_f$ and can become lower than the lowest momentum point 
$p_{min}$ on the torus. Consequently one then sees a pure power law 
in $Z_f(p^2)$ for momenta $p_{min} < p < \alpha$. The critical number
of fermion flavours $N_f^c$ where this transition is observed coincides
with the value of $N_f$ where $Z_f(p_{min}^2)$ is maximal. Thus in a sense
one observes a chiral transition on a torus even for non-vanishing $m_0$ 
when there is none in the infinite volume continuum limit. For the choice 
$m = 0.0024 \, e^2$ the resulting values of $N_f^c$ are plotted against 
$(Le^2)^{1/3}$ in Fig.~\ref{fig:massive}. Apart from our smallest volumes 
the resulting values of $N_f^c$ are significantly larger than in the 
`chiral' limit. We also observe that in the infinite volume limit the critical
number of flavours goes to infinity, in agreement with our findings discussed
in the previous paragraph. Away from the chiral limit QED$_3$ on a torus shows
a chiral phase transition when there is none in the infinite volume/continuum 
limit. This implies that extreme care is needed when one attempts to extract 
information on $N_f^c$ from lattice calculations with finite fermion masses.

\begin{figure}[t]
\epsfig{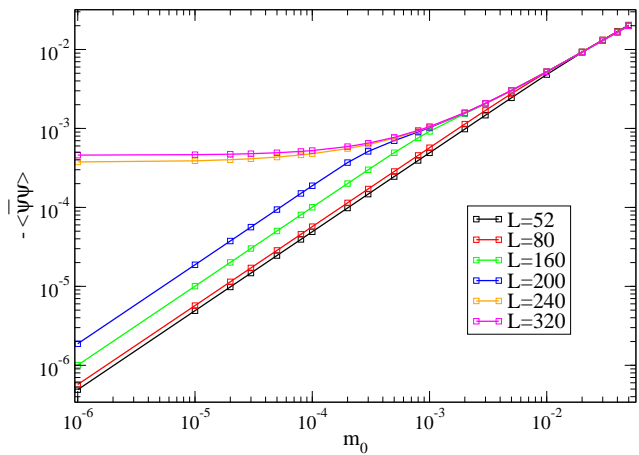}
\caption{The chiral condensate as a function of the fermion mass for a
range of volumes at $N_f = 1.5$. Small volumes give rise to a
vanishing chiral condensate and are nearly degenerate, falsely indicating
weak finite volume effects.}
\label{fig:condensate15}
\end{figure}

Nevertheless it is interesting to compare to lattice results. As explained above the
important scale in assessing finite volume effects is the lowest momentum point
available on the torus compared with the generated fermion mass. In our setup
with antiperiodic boundary conditions in space and time 
directions this scale is given by $L_{DSE} = \sqrt{3} \pi/p_{min}$. 
On the lattice one usually implements antiperiodic boundary conditions in the
temporal direction and periodic ones in the two spatial directions. This
results in $L_{latt} = \pi/p_{min}$ on the lattice. The couplings are related by
$e^2 \simeq 1/(\beta a)$, where $\beta$ is the dimensionless inverse coupling on the
lattice and $a$ the lattice spacing. With $e^2 \simeq 1$ and $\beta \sim \cO(1)$
one then obtains $L_{DSE} e^2 \simeq \sqrt{3} L_{latt} \beta$. This means that
we should compare the results of contemporary lattice calculations on $80^3$-lattices
with our values for $L \approx 140$. From the plot of Fig.~\ref{fig:massive}
we then find chiral symmetry breaking for $N_f=1$, whereas at $N_f=1.5$ the system
is in the chirally symmetric phase. This is in agreement with recent lattice 
results \cite{Strouthos:2008hs} and shows that these results are compatible with
our value of $N_f^c \approx 3.8$ in the infinite volume/continuum limit.

Finally we investigate the behaviour of the chiral condensate on a torus as a function
of the explicit fermion mass. To this end we extract the condensate from our fermion
mass function with the help of Eq.~(\ref{trace-condensate}) at a given fixed 
ultraviolet momentum cutoff. Our results for $N_f=1.5$ and different box lengths $L$ 
are shown in Fig.~\ref{fig:condensate15}. Clearly, for small volumes the condensate
decreases linearly with decreasing bare fermion mass $m_0$ and extrapolates to zero
in the chiral limit. For large enough boxes this behaviour changes and we find
a finite value in this limit in agreement with results from our continuum DSEs. It is
interesting to note that the curves for small volumes are almost degenerate. A volume
analysis in this region would therefore indicate weak finite volume effects where in fact
there are large effects when the volume is increase further. Again, this result should
serve as a caveat for the interpretation of lattice results. 

\section{Summary \label{sec5}}

In this work we investigated volume effects on the chiral phase transition 
of QED$_3$ as a function of the number of flavours, $N_f$. To this end we 
solved a coupled system of Dyson-Schwinger equations for the fermion and 
photon dressing functions in the infinite volume/continuum limit and on a 
three-torus. We worked in a truncation scheme that in the 
infinite volume/continuum limit reproduces a critical number of flavours 
$N_f^c \approx 3.61-3.84$, a number close to the one obtained with more involved
approximation schemes \cite{Fischer:2004nq}. 

Examining the same system on a torus we found considerable volume effects. 
These can be explained due to the presence of scales of vastly different 
magnitude in QED$_3$. On the one hand one has the natural scale  
$\alpha/e^2 = N_f/8$, which is of order one. On the other hand, the 
dynamically generated fermion masses are orders of magnitude smaller.
For example, one has $M(0)=0.05 e^2$, for $N_f=1$ as can be seen from Fig.~\ref{fig:IR}.
While the first scale remains of the same order, the second one rapidly decreases
when the number of flavours becomes larger. When the generated fermion mass
drops below the infrared cutoff of the system, given by the inverse of the 
box length, chiral symmetry breaking disappears and the system falls into the
chirally symmetric phase. We quantified these effects and presented results for
the critical number of flavours as a function of the box length. They agree
qualitatively with corresponding results of~\cite{Gusynin:2003ww}, determined 
in a simpler truncation scheme and a different set-up (continuum limit with 
finite infrared cutoff). Quantitative differences are small. 

Our results 
confirm the notion that lattice calculations at $N_f=1.5$ or $N_f=2$ need 
very large physical volumes to see dynamical chiral symmetry breaking,
let alone quantify their finite volume effects.
Contemporary lattice results cannot yet accommodate for these and consequently
find a system in the chirally symmetric phase in quantitative agreement 
with our findings. We have shown that these results are nicely compatible
with a critical number of flavours $N_f^c \approx 3.61-3.84$ in the infinite
volume/continuum limit.

\section*{Acknowledgement}
We are grateful to Lambert Alff and Jan~M.~Pawlowski for inspiring discussions. 
This work has been supported by a Helmholtz-University Young Investigator 
Grant No. VH-NG-332.



\end{document}